\overfullrule=0pt
\input harvmac
\def\a{{\alpha}}
\def\l{{\lambda}}
\def\b{{\beta}}
\def\g{{\gamma}}
\def\d{{\delta}}

\def\half{{1\over 2}}
\def\p{{\partial}}

\def\t{{\theta}}

\Title{\vbox{\hbox{IFT-P.041/2000 }}}
{\vbox{
\centerline{\bf Consistency of Super-Poincar\'e Covariant
Superstring Tree Amplitudes}}}
\bigskip\centerline{Nathan Berkovits\foot{e-mail: nberkovi@ift.unesp.br}}
\bigskip
\centerline{Brenno Carlini Vallilo\foot{e-mail: vallilo@ift.unesp.br}}
\bigskip
\centerline{\it Instituto de F\'\i sica Te\'orica, Universidade Estadual
Paulista}
\centerline{\it Rua Pamplona 145, 01405-900, S\~ao Paulo, SP, Brasil}

\vskip .3in
Using pure spinors, the superstring was recently 
quantized in a manifestly ten-dimensional
super-Poincar\'e covariant manner and a covariant prescription
was given for tree-level scattering amplitudes. In this paper, 
we prove that this prescription is cyclically symmetric and, for
the scattering of an arbitrary number of massless bosons
and up to four massless fermions, it agrees 
with the standard Ramond-Neveu-Schwarz prescription.

\Date {April 2000}

\newsec{Introduction}

In a recent paper \ref\recent{N. Berkovits, {\it Super-Poincar\'e Covariant
Quantization of the Superstring}, ~~hep-th/0001035.}, a new formalism
was proposed for quantizing the superstring in a manifestly 
ten-dimensional super-Poincar\'e covariant manner. Unlike all previous
such proposals, an explicit covariant prescription was given for computing
tree-level scattering amplitudes of an arbitrary number of
states. 
To check consistency of the formalism,
one would obviously like to prove that this new prescription
for tree amplitudes
is equivalent to
the standard Ramond-Neveu-Schwarz (RNS) prescription\ref\fms{
D. Friedan, E. Martinec and S. Shenker,
{\it Conformal Invariance, Supersymmetry and String Theory},
Nucl. Phys. B271 (1986) 93.}. 

In this paper, we prove this equivalence for tree amplitudes involving
an arbitrary number of massless bosons and up to four massless fermions.
We do not yet have an equivalence proof for 
amplitudes involving massive states or more than four massless fermions,
however, we suspect it might be possible to construct such a proof
using factorization arguments together with the results of this paper.

After reviewing the 
super-Poincar\'e covariant formalism in section 2, we prove in
section 3 that the covariant amplitude prescription for tree
amplitudes is cyclically symmetric, i.e. it does not depend
on which three of the vertex operators are chosen to be unintegrated.
The proof of cyclic symmetry is not the standard one since there
is no natural $b$ ghost in the covariant formalism. In section 4,
we prove by explicit analysis that the covariant and RNS
prescriptions are equivalent for tree amplitudes
involving
an arbitrary number of massless bosons and four massless fermions.
In section 5, we similarly prove equivalence
for tree amplitudes
involving
an arbitrary number of massless bosons and two massless fermions.
And in section 6, we use supersymmetry together with the results of
section 5 to prove
equivalence
for tree amplitudes
involving
an arbitrary number of massless bosons and zero fermions.

\newsec{Review of Super-Poincar\'e Covariant Formalism}

The worldsheet variables in the new formalism include the usual 
ten-dimensional superspace variables $x^m$ and $\t^\a$ ($m=0$ to 9 and
$\a=1$ to 16),
as well as a bosonic spinor variable $\l^\a$
satisfying the pure spinor constraint
$\l^\a \gamma^m_{\a\b} \l^\b=0$
for $m=0$ to 9.\foot{$\g^m_{\a\b}$ and
$\g^{m\,\a\b}$ are $16\times 16$ symmetric matrices which are
the off-diagonal elements of the $32\times 32$ gamma-matrices
and which satisfy
$\g^m_{\a\b} \g^{n\,\b\g}+
\g^n_{\a\b} \g^{m\,\b\g}=2\eta^{mn}\d_\a^\g$ and
$\g_{m\,(\a\b} \g^m_{\g)\d}=0.$}
Although one can solve the pure spinor constraint in terms
of independent variables as in \recent,
this will not be necessary for computing scattering amplitudes. In addition
to the above worldsheet spin-zero variables, the formalism also contains the
worldsheet spin-one variables
$p_\a$ and $N^{mn}$, which
are respectively the conjugate momentum to $\t^\a$ and
the Lorentz currents for $\l^\a$.

The OPE's of these worldsheet variables are
\eqn\ope{p_\a(y) \t^\b(z) \to {\d_\a^\b\over{y-z}},
\quad x^m(y) x^n(z)\to -\eta^{mn}
\log|y-z|,}
$$
N^{mn}(y) \l^\a(z) \to {{(\g^{mn})^\a{}_\b \l^\b(z)}\over {2(y-z)}}, $$
\eqn\nope{N^{kl}(y) N^{mn}(z) \to
{{\eta^{m[l} N^{k]n}(z) - 
\eta^{n[l} N^{k]m}(z) }\over {y-z}} - 3
{{\eta^{kn} \eta^{lm} -
\eta^{km} \eta^{ln}}\over{(y-z)^2}}  ,}
where $(\g^{mn})^\a{}_\b =$
$-(\g^{mn})_\b{}^\a =$
$ \half (\g^{m\,\a\g} \g^n_{\g\b} -
\g^{n\,\a\g} \g^m_{\g\b}) $.
As in 
\ref\csm
{W. Siegel, {\it Classical Superstring Mechanics}, Nucl. Phys. B263 (1986)
93.}, it is convenient to define the combinations
\eqn\defd{
d_\a =p_\a -\half \g^m_{\a\b} \t^\b\p x^m -{1\over 8}\g^m_{\a\b}\g_{m\,\g\d}
\t^\b\t^\g\p\t^\d,\quad
\Pi^m = \p x^m +\half \g^m_{\a\b} \t^\a\p\t^\b} 
which satisfy the OPE's
\eqn\dope{d_\a(y) d_\b(z)\to -{{\g^m_{\a\b}\Pi_m(z)}\over{y-z}},\quad
d_\a(y) \Pi^m(z) \to {{\g^m_{\a\b}\p\t^\b(z)}\over{y-z}},}
and which commute with the spacetime-supersymmetry generator\foot
{We have chosen conventions such that $\{q_\a,q_\b\}=
\g^m_{a\b}\oint dz \p x_m$ to simplify the comparison with RNS amplitudes.}
\eqn\defsusy{q_\a = -\oint dz (p_\a +\half
\g^m_{\a\b} \t^\b\p x^m +{1\over {24}}\g^m_{\a\b}\g_{m\,\g\d}
\t^\b\t^\g\p\t^\d).}

Physical vertex operators in the super-Poincar\'e covariant formalism
are defined to be in the cohomology of the BRST-like operator
$Q=\oint dz \l^\a d_\a$. As discussed in \recent, the unintegrated
massless vertex operator for the open superstring
is 
$U=\l^\a A_\a (x,\t)$ where $A_\a(x,\t)$ is the spinor potential
for super-Yang-Mills satisfying $D_\a (\g^{mnpqr})^{\a\b} A_\b =0$
for any five-form $mnpqr$ and $D_\a ={\partial\over{\partial\t^\a}}
+\half\g^m_{\a\b}\t^\b \p_m$. The spinor potential is defined up to
a gauge transformation $A_\a \to A_\a + D_\a \Omega$, which allows
one to choose the gauge
\eqn\gauge{A_\a(x,\t) = \half a_m(x) \g^m_{\a\b}\t^\b + {1\over 3}
\xi^\g(x) \g^m_{\a\b}\g_{m\,\g\d}\t^\b\t^\d + ...}
where $a_m(x)$ and $\xi^\a(x)$ are 
the linearized on-shell gluon and gluino of
super-Yang-Mills satisfying
$\p^m\p_m a_n =\p^n a_n =\g^m_{\a\b}\p_m\xi^\b=0$
and $...$ denotes terms higher order in $\t$
which depend on derivatives of $a_m$ and $\xi^\a$.
So in the gauge of \gauge, the unintegrated gluon and gluino vertex 
operators are
\eqn\vub{U_m^B = [\half\l\g_m\t +  ...] 
e^{ik\cdot x},\quad 
U_\a^F = [{1\over 3}(\l\g_m\t) (\g^m\t)_\a + ...] e^{ik\cdot x}.}

To compute scattering amplitudes, one also needs to define vertex operators
in integrated form. Although there is no natural $b$ ghost in
this formalism, one can define the integrated vertex operator for
a physical state, 
$U=\int dz V$, by requiring that $[Q,V]=\p U$ where $U$ is the 
unintegrated vertex operator \ref\sieg{W. Siegel,
private communication\semi N. Berkovits, M.T. Hatsuda and
W. Siegel, {\it The Big Picture}, Nucl. Phys. B371 (1992) 434, 
hep-th/9108021.}.
For the massless states, 
$V= \Pi^m A_m + \p \t^\a A_\a + d_\a W^\a + \half N^{mn}F_{mn}$
where $A_m = {1\over 8}\g_m^{\a\b} D_\a A_\b$ is the vector potential,
$W^\a={1\over{10}}\g^{m\,\a\b}
( D_\b A_m - \p_m A_\b)$ is the spinor field strength, and
$F_{mn}= \p_{[m} A_{n]} = {1\over 8}(\g_{mn})^\b{}_\a D_\b W^\a$ is the
vector field strength. So in the gauge of \gauge, the
integrated gluon and gluino vertex 
operators are
\eqn\vib{V_m^B = [\p x_m - ik^n(N_{mn}-\half p\g_{mn}\t) + ...] 
e^{ik\cdot x},\quad 
V_\a^F = -[p_\a + ...] e^{ik\cdot x},}
where the term proportional to $p\g_{mn}\t$ in $V_m^B$ comes from the
$d_\a W^\a$ term in $V$. As will be shown later, the higher-order
$\t$ terms denoted 
by $...$ in \vub\ and \vib\ will not contribute
to tree-level scattering amplitudes involving up to four fermions.

The $N$-point
tree-level scattering amplitude is defined by taking the worldsheet
correlation function of three unintegrated vertex operators and
$N-3$ integrated vertex operators, i.e.
\eqn\presc{{\cal A}=\langle U_1(z_1) U_2(z_2) U_3(z_3) \int dz_4 V_4(z_4) ...
\int dz_N V_N(z_N)\rangle.}
The only subtle point in computing this
correlation function comes from 
the zero modes of $\l^\a$ and $\t^\a$.
The correlation function over these zero modes is defined to vanish
unless one has three $\l$ zero modes and five $\t$ zero modes contracted
in the combination
$(\l\g^m\t) (\l\g^n\t) (\l\g^p\t)(\t \g_{mnp}\t)$. More explicitly, after
performing the correlation function over $x^m$ and over the non-zero
modes of $\t^\a$ and $\l^\a$, the amplitude is obtained by defining
\eqn\norma{
\langle (\l\g^m\t) (\l\g^n\t) (\l\g^p\t)(\t \g_{mnp}\t)\rangle =2880}
where the normalization factor of $2880$ has been chosen to give
agreement with the RNS normalization.
This is equivalent to defining the correlation function over the zero
modes of $Y(x,\t,\l)$ to be proportional to
\eqn\harm{\int d^{10}x \int d\Omega (\bar\l_\rho \l^\rho)^{-3}
\g_{mnp}^{\a\b}
(\bar\l \g^m)^\g
(\bar\l \g^n)^\d
(\bar\l \g^p)^\kappa
\int d\t_\a d\t_\b d\t_\g d\t_\d d\t_\kappa Y,}
where $\bar\l_\a$ is the complex conjugate of $\l^\a$ (after Wick-rotating
to Euclidean space) and $d\Omega$ is an integration over the
different possible orientations of $\l^\a$.
\harm\ can 
be interpreted as integration over an on-shell harmonic superspace since,
as was shown in \recent, it preserves
spacetime-supersymmetry and gauge invariance. Note that integration
over all sixteen $\t$'s leads to inconsistencies as was noted in
\ref\mik{A.R. Mikovic, C.R. Preitschopf and A.E. van de Ven,
{\it Covariant Vertex Operators for the Siegel Superstring}, Nucl. Phys.
B321 (1989) 121.}.

\newsec{Cyclic Symmetry of Tree Amplitudes}

The amplitude prescription of \presc\ fixes three of the vertex operators
to be unintegrated and the remaining vertex operators to be integrated.
The choice of which three vertex operators are unintegrated breaks
the manifest cyclic symmetry of the computation, i.e. the symmetry 
under a cyclic permutation of
the external states. To show that the resulting
amplitude is indeed cyclically symmetric,
one therefore needs to prove that the prescription is independent of
which three vertex operators are chosen to be unintegrated.

In the RNS (or bosonic string) amplitude prescription, the independence
of the choice of which three vertex operators are unintegrated can be proven
using manipulations of the $b$ ghost\ref\polch{J. Polchinski,
{\bf String Theory, Vol. 1}, Cambridge University Press (1998).}.
This follows from the fact that
the integrated vertex operator $\int dz V$ is related to the unintegrated
vertex operator $U$ by $V=\{b,U\}$. In the super-Poincar\'e covariant
formalism, there is no natural candidate for the $b$ ghost so such a proof
cannot be used. 

Alhough one cannot use that $V=\{b,U\}$ in the covariant formalism,
one can use that $[Q, V] = \p U$ \sieg. Note that $[Q,V]=\p U$ is also
satisfied in the RNS and bosonic string, so the proof in this
section serves as an alternative to the conventional proof using
manipulations of the $b$ ghost. Our proof will argue that 
\eqn\proo{\langle U_1(z_1) U_2(z_2) U_3(z_3) \int_{z_3}^{z_1} 
dz_4 V_4(z_4) 
\int_{z_4}^{z_1} dz_5 V_5(z_5) ...
\int_{z_{N-1}}^{z_1} dz_N V_N(z_N)\rangle =}
$$
\langle U_1(z_1) U_2(z_2) \int_{z_2}^{z_3} dy V_3(y) ~U_4(z_3) 
\int_{z_3}^{z_1} dz_5 V_5(z_5) 
\int_{z_5}^{z_1} dz_6 V_6(z_6) 
...
\int_{z_{N-1}}^{z_1} dz_N V_N(z_N)\rangle $$
where $z_1<z_2< ...< z_N$ and the integration with upper limit $z_1$
signifies an integration on the compactified real line which includes the
point at $\infty$. Similar arguments can be used to prove equivalence
of the amplitude prescription for any choice of the three
unintegrated vertex operators.

To prove \proo, first write the left-hand side of \proo\ as
\eqn\prootwo{\langle U_1(z_1) U_2(z_2) \int_{z_2}^{z_3} dy [Q,V_3(y)] 
\int_{z_3}^{z_1} dz_4 V_4(z_4) 
\int_{z_4}^{z_1} dz_5 V_5(z_5) ...
\int_{z_{N-1}}^{z_1} dz_N V_N(z_N)\rangle }
where we have used that 
$\int_{z_2}^{z_3} dy[Q,V_3(y)] =U_3(z_3)-U_3(z_2)$. The contribution coming
from $U_3(z_2)$ can be ignored since when $k_2\cdot k_3$ is sufficiently large,
$U_2(z_2) U_3(z_2 +\epsilon) \to\epsilon^{k_2\cdot k_3}\to 0$ as
$\epsilon\to 0$. But since the amplitude is analytic (except for poles)
in the momentum, 
the contribution coming from $U_3(z_2)$
must vanish for all $k_2$ and $k_3$ if it vanishes
for some region of $k_2$ and $k_3$. This is the `cancelled propagator'
argument discussed in \polch.

Using properties of the correlation function discussed in \recent, one
can pull the BRST operator off of $V_3(z_3)$ until it circles 
either $V_4(z_4)$, 
$V_5(z_5)$, ... , 
$V_N(z_N)$. The contribution coming from when it circles
$V_4(z_4)$ is  
$$\langle U_1(z_1) U_2(z_2) \int_{z_2}^{z_3} dy V_3(y) 
\int_{z_3}^{z_1} dz_4 [-Q,V_4(z_4)]  
\int_{z_4}^{z_1} dz_5 V_5(z_5) ...
\int_{z_{N-1}}^{z_1} dz_N V_N(z_N)\rangle =$$
\eqn\prooth{\langle
U_1(z_1) U_2(z_2) \int_{z_2}^{z_3} dy V_3(y) 
~U_4(z_3)
\int_{z_3}^{z_1} dz_5 V_5(z_5) 
\int_{z_5}^{z_1} dz_6 V_6(z_6) 
...
\int_{z_{N-1}}^{z_1} dz_N V_N(z_N)\rangle }
where the contribution from $U_4(z_1)$ in \prooth\ has been ignored using
the cancelled propagator argument described above. Similarly,
the contributions from $Q$ circling any of $V_5(z_5) ... V_N(z_N)$ can be
ignored since they only give rise to terms which vanish due to the cancelled
propagator argument. Since \prooth\ is equal to the right-hand side
of  \proo, we have proven our claim.

Similar methods can be used to prove that closed
superstring tree amplitudes are
independent of the choice of which three vertex operators
are unintegrated. For the closed superstring, the unintegrated
vertex operator $U(z,\bar z)$ is related to the integrated
vertex operator $\int d^2 z V(z,\bar z)$ by 
\eqn\closed{\{Q[\bar Q, V]\}= \p\bar\p U}
where $Q$ and $\bar Q$ are the holomorphic and anti-holomorphic
BRST operators.  So the closed superstring tree amplitude
\eqn\cltre{{\cal A} = \langle U_1(z_1,\bar z_1)
U_2(z_2,\bar z_2) U_3(z_3,\bar z_3)
\int d^2 z_4 V_4(z_4,\bar z_4) ...
\int d^2 z_N V_N(z_N,\bar z_N)\rangle}
can be written as 
\eqn\cltwo{{\cal A} = {1\over{2\pi}}\langle U_1(z_1,\bar z_1)
U_2(z_2,\bar z_2) \int d^2 y \int d^2 z_4 \log |{{(y-z_3)(z_4-z_3)}
\over{y-z_4}}| \{Q,[\bar Q,V_3(y,\bar y)]\} }
$$  V_4(z_4,\bar z_4) 
\int d^2 z_5 V_5(z_5,\bar z_5) ...
\int d^2 z_N V_N(z_N,\bar z_N)\rangle $$
where we have used that
\eqn\logd{{1\over{2\pi}}\p_y\bar\p_{\bar y}\log
|{{(y-z_3)(z_4-z_3)}
\over{y-z_4}}| = \d^2(y-z_3)- \d^2(y-z_4) }
and that the contribution from $V_3(z_4,\bar z_4)$ can be ignored using the
cancelled propagator argument. Note that the 
argument of the 
logarithm has been chosen such that the logarithm is 
non-singular as $y\to\infty$.
Pulling $Q$ and $\bar Q$ off of 
$V_3(y,\bar y)$, the only contribution comes when they circle
$V_4(z_4,\bar z_4)$
to give 
\eqn\clthr{{\cal A} = {1\over{2\pi}}\langle U_1(z_1,\bar z_1)
U_2(z_2,\bar z_2) \int d^2 y V_3(y,\bar y)
\int d^2 z_4 \log |{{(y-z_3)(z_4-z_3)}
\over{y-z_4}}| }
$$\{Q,[\bar Q,V_4(z_4,\bar z_4)]\}   
\int d^2 z_5 V_5(z_5,\bar z_5) ...
\int d^2 z_N V_N(z_N,\bar z_N)\rangle $$
\eqn\clfour{ = \langle U_1(z_1,\bar z_1)
U_2(z_2,\bar z_2) \int d^2 y V_3(y,\bar y)
V_4(z_3,\bar z_3)   
\int d^2 z_5 V_5(z_5,\bar z_5) ...
\int d^2 z_N V_N(z_N,\bar z_N)\rangle}
which is the closed tree amplitude
prescription with a different choice of unintegrated vertex operators.

It will now be proven that for amplitudes involving an arbitrary
number of massless bosons and up to four massless fermions, the
prescription given by \vub, \vib\ and \presc\ coincides with the
standard RNS prescription of \fms. This will first be proven 
for
amplitudes involving four fermions, then for amplitudes involving
two fermions, and finally, for amplitudes involving zero fermions.

\newsec{Equivalence for Amplitudes involving Four Fermions}

Because of the cyclic symmetry proven in the previous section, one is free
to
choose three of the four fermion vertex operators to be unintegrated.
With this choice, the amplitude prescription of \presc\ is 
\eqn\fourpresc
{{\cal A}=
\langle \xi_1^\a U_\a^F(z_1) \xi_2^\b U_\b^F(z_2)\xi_3^\g  
U^F_\g (z_3)}
$$ \int dz_4 \xi_4^\d V^F_\d(z_4) 
\int dz_5 a_5^m V^B_m(z_5) ...
\int dz_N a_N^n V^B_n(z_N) \rangle $$
where $\xi^\a$ and $a^m$ are the polarizations and
$(U^F_\a,V^F_\a,V^B_m)$ are defined in \vub\ and \vib.
Since $U_\a^F$ has a minimum of two $\t$'s and since
${\cal A}$ requires precisely five $\t$ zero modes to be non-vanishing,
the only terms in
$(U^F_\a,V^F_\a,V^B_m)$ which contribute are
\eqn\fouronly
{{\cal A}=
-{1\over{27}}\langle 
\xi_1^\a f_\a(z_1) \xi_2^\b f_\b(z_2)\xi_3^\g  
f_\g (z_3) \int dz_4 \xi_4^\d p_\d(z_4) }
$$
\int dz_5 a_5^m (\p x_m(z_5) - i k_5^p M_{mp}(z_5)) ...
\int dz_N a_N^n (\p x_n(z_N) - i k_N^q M_{nq}(z_N))
e^{i\sum_{r=1}^N k_r\cdot x(z_r)}
\rangle $$
where $f_\a \equiv(\l\g^m\t)(\g_m\t)_\a$ and
$M_{mn}\equiv N_{mn}-\half(p\g_{mn}\t)$.

The amplitude prescription of 
\fouronly\ will now be shown to coincide with the RNS prescription
of \fms\ with the four fermion vertex operators in the $-\half$
picture.\foot{Comparison of the two prescriptions is 
complicated for amplitudes involving more than four fermions
since such amplitudes require fermion vertex operators in the
$+\half$ picture.}
Choosing three of the fermion vertex operators to be unintegrated,
\eqn\fourrns
{{\cal A}_{RNS}= -\langle 
\xi_1^\a c e^{-{\phi\over 2}}
\Sigma_\a(z_1) \xi_2^\b 
c e^{-{\phi\over 2}}
\Sigma_\b(z_2)  
\xi_3^\g  
c e^{-{\phi\over 2}}
\Sigma_\g(z_3)  
 \int dz_4 \xi_4^\d e^{-{\phi\over 2}}\Sigma_\d(z_4) }
$$
\int dz_5 a_5^m (\p x_m(z_5) - i k_5^p \psi_m\psi_p(z_5)) ...
\int dz_N a_N^n (\p x_n(z_N) - i k_N^q \psi_n\psi_q(z_N)) 
e^{i\sum_{r=1}^N k_r\cdot
x(z_r)}
\rangle $$
where $\Sigma_\a$ is the RNS spin field, 
$-\xi^\a c e^{-{\phi\over 2}}
\Sigma_\a$ is the unintegrated fermion vertex operator, and
$\xi^\a e^{-{\phi\over 2}}
\Sigma_\a$
$=\{b,
-\xi^\a c e^{-{\phi\over 2}}
\Sigma_\a\}$ is the integrated fermion vertex operator.

The correlation function of $x^m$ is clearly equivalent in ${\cal A}$
and ${\cal A}_{RNS}$ of \fouronly\ and \fourrns. So to show 
${\cal A}=
{\cal A}_{RNS}$, one only needs to show that
\eqn\onlyneed{ {1\over{27}}
\langle 
f_\a(z_1)  f_\b(z_2)
f_\g (z_3)p_\d(z_4) 
M_{mp}(z_5) ...
M_{nq}(z_N)\rangle =}
$$
\langle  c e^{-{\phi\over 2}}
\Sigma_\a(z_1) 
c e^{-{\phi\over 2}}
\Sigma_\b(z_2)  
c e^{-{\phi\over 2}}
\Sigma_\g(z_3)  
e^{-{\phi\over 2}}\Sigma_\d(z_4) 
\psi_m\psi_p(z_5) ...
\psi_n\psi_q(z_N) \rangle. $$
To prove \onlyneed,
first note that 
\ope\ implies that \recent
\eqn\Mope{
M_{kl}(y) M_{mn}(z) \to
{{\eta_{m[l} M_{k]n}(z) - 
\eta_{n[l} M_{k]m}(z) }\over {y-z}} + 
{{\eta_{kn} \eta_{lm} -
\eta_{km} \eta_{ln}}\over{(y-z)^2}}  ,}
which coincides with the OPE
of 
$\psi_k\psi_l(y)$ with
$\psi_m\psi_n(z)$. Furthermore, 
\eqn\spinorope{
M_{mn}(y) f_\a(z) \to {{(\g_{mn})_\a{}^\b f_\b(z)}\over {2(y-z)}},\quad 
M_{mn}(y) p_\a(z) \to {{(\g_{mn})_\a{}^\b p_\b(z)}\over {2(y-z)}}, }
reproduces the OPE of $\psi_m\psi_n(y)$ with $\Sigma_\a(z)$.
Since the dependence of ${\cal A}$ and ${\cal A}_{RNS}$ on $z_5 ... z_N$ 
is completely determined by these OPE's, 
we have shown that 
${\cal A}=
{\cal A}_{RNS}$ if 
\eqn\iff{{1\over{27}}
\langle 
f_\a(z_1)  f_\b(z_2)
f_\g (z_3)p_\d(z_4) \rangle =
\langle  c e^{-{\phi\over 2}}
\Sigma_\a(z_1) 
c e^{-{\phi\over 2}}
\Sigma_\b(z_2)  
c e^{-{\phi\over 2}}
\Sigma_\g(z_3)  
e^{-{\phi\over 2}}\Sigma_\d(z_4)\rangle .}

Using the OPE's of \fms, the
right-hand side of \iff\ is easily evaluated to be
\eqn\evalu
{{{\g_{\a\d}^m \g_{m\,\b\g}}\over{z_1-z_4}} + 
{{\g_{\b\d}^m \g_{m\,\g\a}}\over{z_2-z_4}} + 
{{\g_{\g\d}^m \g_{m\,\a\b}}\over{z_3-z_4}}.} 
The left-hand side of \iff\ can also be evaluated by
analyzing the poles of $p_\d(z_4)$. For example, as
$z_4\to z_1$, the left-hand side has a pole whose residue is
\eqn\lhs{ {1\over{27}}\langle ~
[(\g^m_{\a\d}(\l\g_m\t)(z_1) -
(\g_m\l)_\d (\g^m\t)_\a(z_1)] ~ (\l\g^n\t)(\g_n\t)_\b(z_2)
~(\l\g^p\t)(\g_p\t)_\g(z_3)~
 \rangle.}
To simplify the evaluation of \lhs, use
the fact that 
\eqn\simplif{-(\g_m\l)_\d (\g^m\t)_\a = 
\half
[(\g_m\l)_\a (\g^m\t)_\d -
(\g_m\l)_\d (\g^m\t)_\a ] + \half\g^m_{\a\d} (\t\g_m\l)}
$$= Q[\half(\g_m\t)_\a (\g^m\t)_\d] + \half\g^m_{\a\d}(\t\g_m\l)$$
where $Q=\oint dz \l^a d_a.$ Since $Q$ anti-commutes with the 
vertex operators at $z_2$ and $z_3$ and since $\langle Q(Y)\rangle=0$
for any $Y$
\recent, the term 
$-(\g_m\l)_\d (\g^m\t)_\a $ in \lhs\ can be replaced with
$\half\g^m_{\a\d} (\t\g_m\l)$. 
So using the zero mode correlation function defined in \harm,
\lhs\ is equal to ${1\over{18}}\g^m_{\a\d} H_{m\,\b\g}$ where
\eqn\lhstwo{H_{m\,\b\g} =\langle ~
(\l\g_m\t)(z_1)~ (\l\g^n\t)(\g_n\t)_\b(z_2)
~(\l\g^p\t)(\g_p\t)_\g(z_3)~\rangle = 18 \g_{m\,\b\g}.}
To prove \lhstwo, we have used that \harm\ is Lorentz-invariant
so $H_{m\,\b\g}$ must be
proportional to $\g_{m\,\b\g}$. To find the proportionality constant,
we have used from \norma\ that
$$\g^{m\,\b\g}H_{m\,\b\g}=
\langle 
(\l\g_m\t)(\l\g^n\t)
(\l\g^p\t)(\t\g_{mnp}\t)\rangle =2880.$$

So the residue
of the ${1\over{z_1-z_4}}$ pole in \lhs\ is
$\g^m_{\a\d}\g_{m\,\b\g}$ which agrees with
the residue in \evalu. Similarly, one
can show that the residues of the 
${1\over{z_2-z_4}}$ and
${1\over{z_3-z_4}}$ poles agree in the two expressions so we have 
proven that ${\cal A}={\cal A}_{RNS}$ for amplitudes involving four
fermions.

\newsec{Equivalence for Amplitudes involving Two Fermions}

The proof of equivalence for amplitudes involving two fermions
closely resembles the proof for amplitudes involving four fermions.
Choosing two fermion vertex operators and one
boson vertex operator to be unintegrated,
the amplitude prescription in the covariant formalism is 
\eqn\twopresc
{{\cal A}=
\langle \xi_1^\a U_\a^F(z_1) \xi_2^\b U_\b^F(z_2)a_3^m 
U^B_m (z_3) 
\int dz_4 a_4^n V^B_n(z_4) ...
\int dz_N a_N^p V^B_p(z_N) \rangle}
where $\xi^\a$ and $a^m$ are the polarizations and
$(U^F_\a,U^B_m,V^B_m)$ are defined in \vub\ and \vib.
Since $U_\a^F$ has a minimum of two $\t$'s,  
$U_m^B$ has a minimum of one $\t$, and  
${\cal A}$ requires precisely five $\t$ zero modes to be non-vanishing,
the only terms in
$(U^F_\a,U^B_m,V^B_m)$ which contribute are
\eqn\twoonly
{{\cal A}= {1\over{18}}
\langle 
\xi_1^\a f_\a(z_1) \xi_2^\b f_\b(z_2)a_3^m  
b_m (z_3) }
$$
\int dz_4 a_4^n (\p x_n(z_4) -i k_4^q M_{nq}(z_4)) ...
\int dz_N a_N^p (\p x_p(z_N) -i k_N^r M_{pr}(z_N))
e^{i\sum_{r=1}^N k_r\cdot
x(z_r)}
\rangle $$
where $f\equiv(\l\g^m\t)(\g_m\t)_\a$, $b_m\equiv \l\g_m\t$, and
$M_{mn}\equiv N_{mn}-\half(p\g_{mn}\t)$.

The amplitude prescription of 
\twoonly\ will now be shown to coincide with the RNS prescription
of \fms,
\eqn\tworns
{{\cal A}_{RNS}=\langle 
\xi_1^\a c e^{-{\phi\over 2}}
\Sigma_\a(z_1) \xi_2^\b 
c e^{-{\phi\over 2}}
\Sigma_\b(z_2)  
a_3^m c e^{-\phi} \psi_m(z_3)  }
$$
\int dz_4 a_4^n (\p x_n(z_4) -i k_4^q \psi_n\psi_q(z_4)) ...
\int dz_N a_N^p (\p x_p(z_N) -i k_N^r \psi_p\psi_r(z_N)) 
e^{i\sum_{r=1}^N k_r\cdot
x(z_r)}
\rangle $$
where the fermion vertex operators are in the $-\half$
picture and the unintegrated boson vertex operator is in the $-1$ picture.

As before, the correlation function of $x^m$ is equivalent in ${\cal A}$
and ${\cal A}_{RNS}$ of \twoonly\ and \tworns. Furthermore,
\eqn\vectorope{
M_{mn}(y) f_\a(z) \to {{(\g_{mn})_\a{}^\b f_\b(z)}\over {2(y-z)}},\quad 
M_{mn}(y) b_p(z) \to {{\eta_{np} b_m(z)-\eta_{mp} b_n(z)}\over {y-z}}, }
reproduces the OPE of $\psi_m\psi_n(y)$ 
with $\Sigma_\a(z)$ and with $\psi_p(z)$.
So using the arguments of the previous section, ${\cal A}=
{\cal A}_{RNS}$ if 
\eqn\twoiff{ {1\over{18}}
\langle 
f_\a(z_1)  f_\b(z_2)
b_m (z_3) \rangle =
\langle  c e^{-{\phi\over 2}}
\Sigma_\a(z_1) 
c e^{-{\phi\over 2}}
\Sigma_\b(z_2)  
c e^{-{\phi}}
\psi_m(z_3)\rangle.}  

Using the RNS OPE's of \fms, the
right-hand side of \twoiff\ is easily evaluated to be $\g_{m\,\a\b}.$
The left-hand side of \iff\ 
is
\eqn\lhs{ {1\over{18}}\langle ~
(\l\g^n\t)(\g_n\t)_\a (z_1)
~(\l\g^p\t)(\g_p\t)_\b(z_2)~
(\l\g_m\t)(z_3)
 \rangle =  {1\over{18}}H_{m\,\a\b} = \g_{m\,\a\b}}
from \lhstwo. So we have 
proven that ${\cal A}={\cal A}_{RNS}$ for amplitudes involving two
fermions.

\newsec{Equivalence for Amplitudes involving Zero Fermions}

The equivalence of amplitudes involving zero fermions
will now be proven using spacetime 
supersymmetry to relate these amplitudes with amplitudes
involving two fermions. This will be made explicit uusing
the supersymmetry transformations of the covariant and RNS
massless vertex operators.

First, note that the supersymmetry generator of \defsusy\ exchanges
the massless boson and fermion vertex operators of \vub\ and
\vib\ in the following manner:
\eqn\exch{\{q_\a, U_m^B\} = {i\over 2}
k^n(\g_{mn})_\a{}^\b U_\b^F +Q(\Omega_{m\a}),\quad 
[q_\a, U_\b^F] = \g^m_{\a\b} U_m^B + Q(\Sigma_{\a\b}),}
$$
[q_\a, V_m^B] = {i\over 2}
 k^n(\g_{mn})_\a{}^\b V_\b^F - \p(\Omega_{m\a}),\quad 
\{q_\a, V_\b^F\} = \g^m_{\a\b} V_m^B + \p(\Sigma_{\a\b}),$$
for some $\Omega_{m\a}$ and $\Sigma_{\a\b}$. \exch\
can be derived either by explicit computation or by using the on-shell
supersymmetry transformations of the super-Yang-Mills component fields.
The dependence on 
$\Omega_{m\a}$ and $\Sigma_{\a\b}$ comes from the fact that
supersymmetry transformations do not commute with the gauge choice of
\gauge.

The covariant
amplitude prescription for the scattering of $N$ massless bosons is
\eqn\boson
{{\cal A}=\langle a_1^m U_m^B(z_1) a_2^n U_n^B(z_2) a_3^p 
U^B_p (z_3) 
\int dz_4 a_4^q V^B_q(z_4) ...
\int dz_N a_N^r V^B_r(z_N) \rangle, }
which can be written using \exch\ and BRST-invariance of the
correlation function as
\eqn\susyboson
{{\cal A}={1\over{16}}\langle a_1^m \g_m^{\a\b}
[q_\a, U_\b^F(z_1)] a_2^n U_n^B(z_2) a_3^p 
U^B_p (z_3) 
\int dz_4 a_4^q V^B_q(z_4) ...
\int dz_N a_N^r V^B_r(z_N) \rangle. }
Since the correlation function preserves supersymmetry as was
shown in \recent, $q_\a$ can be pulled off of $U_\b^F(z_1)$ until
it circles any of the other boson vertex operators. 

For example, when
$q_\a$ circles $U_n^B(z_2)$, one gets the term
\eqn\onesuch{-{1\over{16}}
\langle a_1^m \g_m^{\a\b}
 U_\b^F(z_1) a_2^n \{q_\a,U_n^B(z_2)\} a_3^p 
U^B_p (z_3) 
\int dz_4 a_4^q V^B_q(z_4) ...
\int dz_N a_N^r V^B_r(z_N) \rangle }
$$=-{i\over{32}}
\langle a_1^m \g_m^{\a\b}
 U_\b^F(z_1) a_2^n k_2^s(\g_{ns})_\a{}^\d U_\d^F(z_2) a_3^p 
U^B_p (z_3) 
\int dz_4 a_4^q V^B_q(z_4) ...
\int dz_N a_N^r V^B_r(z_N) \rangle. $$
But using the results of section 5, this is equal to
the analagous RNS correlation function
where $U_\a^F$ is replaced with 
the picture $-\half$ fermion vertex operator
$-c e^{-{\phi\over 2}}\Sigma_\b e^{ik\cdot x}$, 
$U_m^B$ is replaced with the picture $-1$
boson vertex operator 
$c e^{-\phi}\psi_p e^{ik\cdot x}$, and $V_m^B$ is replaced with 
$(\p x_m - i k^n \psi_m\psi_n) e^{ik\cdot x}$.

Similarly, when 
$q_\a$ circles $V_q^B(z_4)$, one gets the term
\eqn\twosuch{-{i\over{32}}
\langle a_1^m \g_m^{\a\b}
 U_\b^F(z_1) a_2^n U_n^B(z_2) a_3^p 
U^B_p (z_3) 
\int dz_4 a_4^q k_4^s(\g_{qs})_\a{}^\d V^F_\d(z_4) }
$$
\int dz_N a_5^r V^B_r(z_5) ...
\int dz_N a_N^s V^B_s(z_N) \rangle .$$
To relate \twosuch\ to an analogous RNS expression,
one first uses the results of section 3 to exchange
$U_p^B(z_3)\int dz_4 V^F_\d(z_4)$ for $\int dy V_p^B(y) ~U^F_\d(z_3)$.
One can then use the results of section 5 to relate \twosuch\ to
an analogous RNS expression as was done for \onesuch.

So ${\cal A}$
of \boson\ is equal to a sum of RNS correlation functions
involving $N-2$ massless boson vertex operators and two massless
fermion vertex operators. It will now be shown that this sum
of RNS correlation functions is related by supersymmetry to
the RNS prescription for the scattering of $N$ massless bosons:
\eqn\bosonrns{{\cal A}_{RNS}=
\langle \{Q,\xi(z_0)\}
a_1^m c e^{-\phi}\psi_m(z_1) a_2^n c e^{-\phi}\psi_n(z_2) a_3^p 
c e^{-\phi}\psi_p (z_3) }
$$
\int dz_4 a_4^q (\p x_q(z_5) - i k_5^s \psi_q\psi_s(z_4)) ...
\int dz_N a_5^r (\p x_r(z_N) - i k_N^t \psi_r\psi_t(z_N)) 
e^{i\sum_{r=1}^N k_r\cdot
x(z_r)}
 \rangle, $$
where 
$\{Q,\xi(z_0)\}$ is the picture-raising operator and $z_0$ is arbitrary.
To prove ${\cal A}={\cal A}_{RNS}$, first write 
\eqn\rnstwo{{\cal A}_{RNS}={1\over{16}}
\langle \{Q,\xi(z_0)\}
a_1^m \g_m^{\a\b} [q^{RNS}_\a, - c e^{-{\phi\over 2}}\Sigma_\b(z_1)]
 a_2^n c e^{-\phi}\psi_n(z_2) a_3^p 
c e^{-\phi}\psi_p (z_3) }
$$
\int dz_4 a_4^q (\p x_q(z_4) - i k_4^s \psi_q\psi_s(z_4)) ...
\int dz_N a_5^r (\p x_r(z_N) - i k_N^t \psi_r\psi_t(z_N)) 
e^{i\sum_{r=1}^N k_r\cdot
x(z_r)}
\rangle, $$
where $q^{RNS}_\a = \oint dz e^{-{\phi\over 2}}\Sigma_\a$ is the
RNS spacetime-supersymmetry generator in the $-\half$ picture\fms.
Pulling $q_\a^{RNS}$ off of 
$-c e^{-{\phi\over 2}}\Sigma_\b(z_1)$ until it circles the other
vertex operators,
one recovers precisely the same terms as found earlier.

For example, if $q_\a^{RNS}$ circles 
$ c e^{-\phi}\psi_n(z_2)$, one obtains the term
$ \g_{n\,\a\b}c e^{-{{3\phi}\over 2}}\Sigma^\b(z_2)$. Choosing
$z_0=z_2$, one gets the picture-raised version of this term which is
${i\over 2}k_2^s (\g_{ns})_\a{}^\d$
$ c e^{-{{\phi}\over 2}}\Sigma_\d(z_2)$. Comparing
expressions, one sees that this is
precisely the RNS version of \onesuch.
Similarly, if 
$q_\a^{RNS}$ circles 
$\p x_q(z_4) - i k_4^s \psi_q\psi_s(z_4) $, one obtains
$ {i\over 2} 
k_4^s (\g_{qs})_\a{}^\b e^{-{{\phi}\over 2}}\Sigma_\b(z_4)$. Choosing
$z_0=z_3$ to convert
$c e^{-\phi} \psi_p(z_3)$ to
$c(\p x_p(z_3)-i k_3^u \psi_p\psi_u(z_3))$
and using cyclic symmetry to exchange the 
integrated and unintegrated vertex operators
at $z_3$ and $z_4$, one recovers the RNS version of \twosuch.
So we have proven the equivalence of amplitudes involving zero fermions. 

{\bf Acknowledgements:} NB would like to thank 
Warren Siegel and Edward Witten
for useful discussions and CNPq grant 300256/94-9
for partial financial support. BCV would like to thank Osvaldo
Chandia for useful discussions and Fapesp
grant 98/15374-1 for financial support.

\listrefs

\end